\documentclass[a4paper,11pt]{article}
\pdfoutput=1 

\usepackage{jcappub} 

\usepackage[T1]{fontenc} 
\usepackage{multirow}
\usepackage{hyperref}
\usepackage{ulem}

\newcommand{\lcdm}{$\Lambda$CDM}
\newcommand{\lcdmrc}{$\Lambda\mathrm{CDM}+r+\chi$ }

\newcommand{\loga}{$\ln[10^{10}A_s]$}

\title{\boldmath Testing chirality of primordial gravitational waves with Planck and future CMB data: no hope from angular power spectra}

 \author[a,b]{Martina~Gerbino,}
 \author[c,d]{Alessandro~Gruppuso,}
 \author[e]{Paolo~Natoli,}
 \author[f]{Maresuke~Shiraishi,}
 \author[g]{Alessandro Melchiorri}

\affiliation[a]{The Oskar Klein Centre for Cosmoparticle Physics, Department of Physics, Stockholm University, AlbaNova, SE-106 91 Stockholm, Sweden}
\affiliation[b]{The Nordic Institute for Theoretical Physics (NORDITA), Roslagstullsbacken 23, SE-106 91 Stockholm, Sweden}
\affiliation[c]{INAF, Istituto di Astrofisica Spaziale e Fisica Cosmica di Bologna, \\
via P.~Gobetti 101, I-40129 Bologna, Italy}
\affiliation[d]{INFN, Sezione di Bologna, Via Irnerio 46, I-40126 Bologna, Italy}
\affiliation[e]{Dipartimento di Fisica e Scienze della Terra and INFN,
Universit\`a degli Studi di Ferrara, Via Saragat 1, I-44100 Ferrara, Italy}
\affiliation[f]{Kavli Institute for the Physics and Mathematics of the Universe (Kavli IPMU, WPI), UTIAS, The University of Tokyo, Chiba, 277-8583, Japan}
\affiliation[g]{Physics Department and INFN, Universit\`a di Roma
``La Sapienza'', P.le\ Aldo Moro 2, 00185, Rome, Italy}

\emailAdd{martina.gerbino@fysik.su.se}
\emailAdd{gruppuso@iasfbo.inaf.it}
\emailAdd{paolo.natoli@gmail.com}
\emailAdd{maresuke.shiraishi@ipmu.jp}
\emailAdd{alessandro.melchiorri@roma1.infn.it}

\abstract{We use the 2015 Planck likelihood in combination with the Bicep2/Keck likelihood (BKP and BK14) to constrain the chirality, $\chi$, of primordial gravitational waves in a scale-invariant scenario. 
In this framework, the parameter $\chi$ enters theory always coupled to the tensor-to-scalar ratio, $r$, e.g. in combination of the form $\chi \cdot r$.
Thus, the capability to detect $\chi$ critically depends on the value of $r$.
We find that with present data sets $\chi$ is \textit{de facto} unconstrained. 
We also provide forecasts for $\chi$ from future CMB experiments, including COrE+, exploring several fiducial values of $r$. 
We find that the current limit on $r$ is tight enough to disfavor a neat detection of $\chi$. For example, in the unlikely case in which $r\sim0.1(0.05)$, the maximal chirality case, i.e. $\chi = \pm1$, could be detected with a significance of $\sim2.5(1.5)\sigma$ at best. We conclude that the two-point statistics at the basis of CMB likelihood functions is currently unable to constrain chirality and may only provide weak limits on $\chi$ in the most optimistic scenarios. Hence, it is crucial to investigate the use of other observables, e.g. provided by higher order statistics, to constrain these kinds of parity violating theories with the CMB.}

\begin{document}
\begin{flushright}
{\small IPMU16-0076}
\end{flushright}
\maketitle
\flushbottom

\section{Introduction}
\label{intro}

Parity symmetry is one of the essential properties of the gravity and electromagnetic sectors in the Universe. It is preserved in the description provided by general relativity and standard Maxwell electromagnetism, while its breaking
might give indication of the deviation from such a standard models. Parity violation may occur within inflationary models through circularly polarized gravitational waves (GWs), which are referred as chiral gravity models, and also at late-time Universe through a new Chern-Simons like coupling via the so called cosmological birefringence effect \cite{Lue:1998mq,Kostelecky:2002hh,Finelli:2008jv,Kahniashvili:2008va}.

Polarized cosmic microwave background (CMB) observations can be used to test parity symmetry. Usual four CMB power spectra of the temperature and E/B-mode polarization anisotropies, i.e., TT, TE, EE, and BB, are always nonzero regardless of parity. On the other hand, additional two combinations, TB and EB, are different from zero when parity is violated, otherwise they are null, and hence unbiased observables of parity violation \cite{Lue:1998mq}. There are constraints on many kinds of chiral gravity models and cosmological birefringence models obtained from many kinds of CMB data \cite{Feng:2006dp,Xia:2007qs,Saito:2007kt,Cabella:2007br,Kahniashvili:2008va,Gubitosi:2009eu,
Das:2009ys,Gruppuso:2011ci,Gluscevic:2012me,Gubitosi:2012rg,Kaufman:2013vbd,Kahniashvili:2014dfa,
Gubitosi:2014cua,Galaverni:2014gca,Ade:2015cva,Ade:2015cao,Gruppuso:2015xza,Aghanim:2016fhp}, indicating no significant evidence of parity violation.

In this paper, we test chiral gravity models with the most recent all-sky polarized data observed by the {\it Planck} satellite. The shapes of the GW spectrum and the resultant TB and EB spectra are strongly model-dependent (e.g., \cite{Lue:1998mq,Alexander:2004wk,Lyth:2005jf,Takahashi:2009wc,Alexander:2009tp,
Satoh:2010ep,Dyda:2012rj,Wang:2012fi,Sorbo:2011rz,Barnaby:2011vw,
Barnaby:2012xt,Ferreira:2014zia,Caprini:2014mja,Bartolo:2014hwa,Bielefeld:2014nza,Ferreira:2015omg,Namba:2015gja,
Obata:2016tmo,Maleknejad:2016qjz,Ben-Dayan:2016iks}), while in this paper, as the simplest example, we constrain an almost scale-invariant template, which has been analyzed in \cite{Saito:2007kt,Gluscevic:2010vv}. This is motivated by the fact that
CMB $\ell$ modes of the TB and EB data used in our analysis are limited to the range $2 \leq \ell \leq 29$
and hence it is hard to extract useful information on the scale dependence. We then consistently restrict ourselves to the scale-independent case even when combining \textit{Planck} data with Bicep/Keck measurements or when perform forecasts for a COrE+ like experiment, as explained below in more detail. The analysis with the WMAP TB and EB data have led to an unconstrained result due to the lack of sensitivity \cite{Saito:2007kt}. On the other hand, according to the Fisher matrix analyses, there could be a region of the parameter space where visibly large TB and EB correlation are produced, if a {\it Planck}-level sensitivity is realized \cite{Saito:2007kt,Gluscevic:2010vv}. This motivates the check with the {\it Planck} data, although we would like to address the fact that forecasts provided in \cite{Saito:2007kt,Gluscevic:2010vv} assume a specific fiducial value for $r$ and rely on the expected Planck-HFI sensitivity in the $143\,\mathrm{GHz}$ channel.

In this paper, we perform a Monte Carlo analysis for deriving updated constraints on the chirality, conveniently parameterized, both employing current CMB data from the \textit{Planck} satellite in combination with Bicep/Keck measurements of the B-modes at degree angular scales and providing forecasts for a future satellite mission like COrE+. 




The paper is organized as follows: in Section \ref{sec:model} we introduce the considered chiral gravity model giving the main equations and defining the additional parameter $\chi$ which basically 
provides the fraction of circularly polarized gravitational wave; in Section \ref{sec:method} we describe the data set considered to provide the constraints on $\chi$ and the other cosmological parameters; 
furthermore, in the same section we provide forecast for future CMB experiments, as COrE+. The latter is performed using both a Fisher matrix or Markov Chain Monte Carlo (MCMC) approach; 
conclusions are drawn in Section \ref{sec:conclusions}.

\section{Chiral gravity model}
\label{sec:model}

The detection of non-vanishing chiral GWs would be a powerful evidence of the Chern-Simons interactions in the very primordial Universe. For example, if there exists an axion or a pseudoscalar field $\phi$ in the inflationary era and it couples to a gauge field $A_\mu$ via an electromagnetic Chern-Simons interaction $f(\phi) \tilde{F}F$, the U(1) gauge field is helical and sources chiral GWs due to the inverse decay process $A + A \rightarrow h$ \cite{Sorbo:2011rz,Barnaby:2011vw,Barnaby:2012xt,Adshead:2013qp,Adshead:2013nka,Caprini:2014mja,Bartolo:2014hwa,Namba:2015gja}. The similar production can be realized also by the SU(2) gauge field \cite{Bielefeld:2014nza,Obata:2016tmo,Maleknejad:2016qjz}. Another well-known candidate is a gravitational Chern-simons term $g(\phi) \tilde{R}R$, motivated by the extention or modification of general relativity. Provided that $g(\phi)$ shows a time dependence, then it is no longer a topological term and can affect the GW production. This term explicitly breaks parity and hence the induced GW becomes chiral \cite{Lue:1998mq,Alexander:2004wk,Lyth:2005jf,Takahashi:2009wc,Alexander:2009tp,Satoh:2010ep,Dyda:2012rj,Wang:2012fi}.

The shape of resultant GW power spectrum is strongly model-dependent. Specific scale dependence can be created e.g., by choosing time dependence of the running coupling, $f(\phi)$ or $g(\phi)$. In the next section, we do the data analysis with a nearly scale-invariant power spectrum template, since our available CMB $\ell$ modes, which are limited to $2 \leq \ell \leq 29$, are too few to extract useful information on the scale dependence. Such a nearly-scale invariant power spectrum is realized e.g., in the simplest pseudoscalar inflation models where $f(\phi) \propto \phi$ with $\phi$ identified with the inflaton field \cite{Sorbo:2011rz,Barnaby:2011vw,Meerburg:2012id,Maleknejad:2016qjz}.

Let us decompose primordial GWs $h_{ij} = \delta g_{ij}^{TT} / a^2$, with $a$ denoting the scale factor, into two helicity states ($\lambda = \pm 2$):
\begin{eqnarray}
h_{ij}(\textbf{x})  = \int\frac{d^{3}{\bf k}}{(2\pi)^{3}}\sum_{\lambda = \pm 2}
h_{\bf k}^{(\lambda)} 
e^{(\lambda)}_{ij}(\hat{\bf k})e^{i{\bf k}\cdot {\bf x}} ~,
\end{eqnarray}
where we have used the transverse-traceless polarization tensor $e_{ij}^{(\pm 2)}$ satisfying $\hat{k}_{i} e_{ij}^{(\lambda)}(\hat{\bf k}) = e_{ii}^{(\lambda)}(\hat{\bf k}) = 0$, $e_{ij}^{(\lambda) *}(\hat{\bf k}) = e_{ij}^{(-\lambda)}(\hat{\bf k}) = e_{ij}^{(\lambda)}(- \hat{\bf k})$ and $e_{ij}^{(\lambda)}(\hat{\bf k}) e_{ij}^{(\lambda')}(\hat{\bf k}) = 2 \delta_{\lambda, -\lambda'}$. Assuming isotropy and homogeneity of the Universe, the GW power spectrum can be expressed as
\begin{eqnarray}
\left\langle h_{{\bf k}_1}^{(\lambda_1)} h_{{\bf k}_2}^{(\lambda_2)} \right\rangle
= (2\pi)^3 \frac{2\pi^2}{k_1^3} \frac{{\cal P}_h^{(\lambda_1)}(k_1)}{2}
\delta^{(3)}({\bf k}_1 + {\bf k}_2)
  \delta_{\lambda_1, \lambda_2} ~.
\end{eqnarray}
In this convention, the GW helicity $\lambda = \pm 2$ is exchanged each other under parity transformation, so parity violation in the power spectrum equates to ${\cal P}_h^{(+2)} \neq {\cal P}_h^{(-2)}$. Following the convention in the previous literature \cite{Saito:2007kt,Gluscevic:2010vv}, we define the tensor-to-scalar ratio $r$ and the chirality parameter $\chi$ as
\begin{eqnarray}
 r &\equiv&  \frac{\left\langle h^{ij}({\bf k}_1)h_{ij}({\bf k}_2)\right\rangle}{\left\langle \zeta({\bf k}_1) \zeta({\bf k}_2) \right\rangle}
 = \frac{{\cal P}_h^{(+2)}(k_1) + {\cal P}_h^{(-2)}(k_1)}{{\cal P}_\zeta(k_1)} ~, \\
 \chi &\equiv& \frac{{\cal P}_h^{(+2)}(k) - {\cal P}_h^{(-2)}(k)}{{\cal P}_h^{(+2)}(k) + {\cal P}_h^{(-2)}(k)}
 = \frac{{\cal P}_h^{(+2)}(k) - {\cal P}_h^{(-2)}(k)}{r {\cal P}_\zeta(k)}
\end{eqnarray}
where ${\cal P}_\zeta(k)$ is the power spectrum of the curvature perturbation $\zeta$, defined in
\begin{eqnarray}
\left\langle \zeta_{{\bf k}_1} \zeta_{{\bf k}_2} \right\rangle
= (2\pi)^3 \frac{2\pi^2}{k_1^3} {\cal P}_\zeta(k_1)
\delta^{(3)}({\bf k}_1 + {\bf k}_2)~.
\end{eqnarray}
Note that $\chi$ takes nonzero values respecting $-1 \leq \chi \leq 1$, if parity is violated.

The CMB temperature ($X = T$) and polarization ($X = E, B$) anisotropy is expanded as $X(\hat{\bf n}) = \sum_{\ell m} a_{\ell m}^X Y_{\ell m}(\hat{\bf n})$, where $\hat{\bf n}$ is the unit vector from the observer to CMB photons. The harmonic coefficients induced by primordial GWs are expressed as \cite{Pritchard:2004qp, Saito:2007kt, Weinberg:2008zzc, Shiraishi:2010sm}
\begin{eqnarray}
a_{\ell m}^{T/E}
&=& 4\pi i^{\ell} \int \frac{d^3 {\bf k}}{(2\pi)^{3}}
 \Delta_{\ell}^{T/E}(k) 
\left( h_{\bf k}^{(+2)} {}_{-2}Y_{\ell m}^*(\hat{\bf k})
+ h_{\bf k}^{(-2)} {}_{2}Y_{\ell m}^*(\hat{\bf k})
\right)  ~,
\\
a_{\ell m}^{B}
&=& 4\pi i^{\ell} \int \frac{d^3 {\bf k}}{(2\pi)^{3}}
 \Delta_{\ell}^B(k) 
\left( h_{\bf k}^{(+2)} {}_{-2}Y_{\ell m}^*(\hat{\bf k}) 
- h_{\bf k}^{(-2)} {}_{2}Y_{\ell m}^*(\hat{\bf k})
\right)  ~,
\label{eq:alm}
\end{eqnarray}
where $\Delta_{\ell}^{T/E/B}(k)$ is the tensor-mode radiation transfer function. Using these, one can formulate the CMB power spectra:
\begin{eqnarray}
\left\langle a_{\ell_1 m_1}^{X_1} a_{\ell_2 m_2}^{X_2} \right\rangle
&=& (-1)^{m_1} \delta_{\ell_1, \ell_2} \delta_{m_1, -m_2} C_{\ell_1}^{X_1 X_2} ~, \\
  C_\ell^{X_1 X_2} &=& 4\pi \int_0^\infty \frac{dk}{k} \Delta_{\ell}^{X_1}(k) \Delta_{\ell}^{X_2}(k) \frac{r {\cal P}_\zeta(k)}{2} \times 
  \begin{cases}
   1 &: X_1 X_2 = TT, EE, BB, TE \\
   \chi &: X_1 X_2 = TB, EB
    \end{cases}
  ~, \label{eq:Cl}
\end{eqnarray}

In the following data analysis, we assume ${\cal P}_h^{(\pm 2)}(k) = {\rm const}$ (since we are interested in the scale-invariant case as explained above), however, we take into account a slightly red-tilted shape of the curvature power spectrum to respect observations.

As seen in eq.~\eqref{eq:Cl}, the TB and EB correlation depend on the combination of $r$ and $\chi$, it is thus essentially difficult to measure $\chi$ from TB and EB if $r$ is very small. Nevertheless, the Fisher matrix forecast assuming a {\it Planck}-level sensitivity\footnote{We recall that previous forecasts like e.g. \cite{Saito:2007kt,Gluscevic:2010vv} rely on Planck-HFI expected sensitivity.} tells that, even if $r \simeq 0.1$, $\chi = 1$ can be judged with $\gtrsim 1 \sigma$ accuracy \cite{Saito:2007kt,Gluscevic:2010vv}. This motivates our analysis with the {\it Planck} data.

\section{Method and datasets}
\label{sec:method}

\subsection{Datasets}
We derive constraints on the model under investigation from current data and perform forecasts for future experiments. 
As our current dataset, we make use of the full set of the Planck 2015 likelihood in both temperature and polarization (referred as Planck TT,TE,EE+lowTEB) \cite{Aghanim:2015xee}. 
Although we expect that the chirality has the main impact on the large-scale region of the spectra for the nearly scale-invariant case under examination, we also employ small-scale data in order to better constrain the remaining cosmological parameters and possibly break degeneracies among them. In particular, we want to reduce the degeneracy between the tensor-to-scalar ratio and the amplitude of scalar perturbations. 
We use the \textit{Planck} likelihood alone or complemented with results from the Bicep/Keck collaboration. 
We test both the data coming from the joint Planck/Bicep/Keck analysis (BKP \cite{Ade:2015tva}) and the most recent results from the Bicep/Keck collaboration (BK14 \cite{Array:2015xqh}). 
In both cases, we include the likelihood accounting for the B auto-correlation modes at degree angular scales, where we expect the recombination bump. We follow the default setting of considering the first five bandpowers
for BKP, roughly corresponding to the multipole range $40<\ell<180$, while BK14 dataset roughly covers the multipole range $40<\ell<300$ in 9 bandpowers. 
The inclusion of BKP and BK14 data is motivated by the fact that it allows to better constrain the tensor-to-scalar ratio $r$, thus reducing any possible degeneracy between the latter parameter and $\chi$. 

As far as forecasts are concerned, we simulate a COrE+ like mission, assuming the experimental setup described in \cite{2015hsa8.conf..334R}, 
which corresponds to the dual-band detector upgrade of the baseline COrE+ proposal.
We suppose a nine-frequency measurement of the CMB signal in the frequency range [90-220] GHz over a fraction $f_{\mathrm{sky}}=0.70$ of the sky, in the best-case scenario of perfect foreground removal. 
We employ the full set of lensed temperature and polarization power spectra up to $\ell=3000$, which is a reasonable range an experiment like COrE+ may achieve. Details about the generation of mock CMB data can be found in \cite{Bond:1998qg,Bond:1997wr}.
We report in Table \ref{tab:fiducial} our fiducial models. We decide to test two different cases, namely the absence of chirality signal $\chi=0$ and maximal chirality signal $\chi=1$, 
for different amplitudes of the primordial tensor modes. As we can see from Table \ref{tab:fiducial}, we choose to consider also models with a tensor-to-scalar ratio which is already excluded 
with high significance by current data (see e.g. \cite{Array:2015xqh} for the most recent results), i.e.  $r\ge0.5$. The reason for this choice is to highlight that the detectability of the chiral models analysed in this work is crucially related 
to the amplitude of the primordial tensor signal, as $C_\ell^{TB}$ and $C_\ell^{EB}$ depend on the combination of $r$ and $\chi$, as described in Eq.~\eqref{eq:Cl}.
The remaining cosmological parameters are set to the Planck TT,TE,EE+lowTEB $\Lambda$CDM best-fit model\footnote{The full grid of Planck results can be found at \url{http://pla.esac.esa.int/pla/}.}. 

\subsection{Monte Carlo analysis}
We perform a Monte Carlo Markov Chain (MCMC) analysis by using the public code \texttt{cosmomc} \cite{cosmomc}, complemented with the Boltzmann solver \texttt{camb}, modifying the relevant subroutines for the inclusion of non-vanishing primordial 
$TB$ and $EB$ spectra. We consider an 8-dimensional parameter space representative of our model. 
The parameter vector is composed by the baryon density $\Omega_bh^2$, the cold dark matter density $\Omega_ch^2$, the angular size of the sound horizon at decoupling $\theta$, 
the reionization optical depth $\tau$, the amplitude $\ln[10^{10}A_s]$ and tilt $n_S$ of the power spectrum of primordial scalar perturbations at a pivot scale of $k=0.05\,\mathrm{Mpc^{-1}}$, 
the tensor-to-scalar ratio $r$ at a pivot scale of $k=0.05\,\mathrm{Mpc^{-1}}$ and the chirality parameter $\chi$. As anticipated in the previous sections, we choose to test chirality within the framework 
of a scale-invariant primordial tensor spectrum, i.e. with a vanishing tensor spectral index $n_T=0$. We note that, given the current upper limit on the $r$, our choice is not dissimilar from having assumed 
the standard inflation consistency relation. We leave to future works the possibility to test chirality in scenarios where consistency relation does not hold.
However, we do not expect significant deviations with respect to the findings reported in this work, given the current limits on $n_T$ \cite{Cabass:2015jwe}.  

When we perform forecasts, we employ an exact likelihood approach for the Monte Carlo analysis (see e.g. \cite{Lewis:2005tp, Perotto:2006rj}).

In Appendix \ref{appen:fish}, we report results from a Fisher matrix approach as a consistency check of our findings from the Monte Carlo analysis.

\begin{table}
\begin{center}
\begin{tabular}{c|c}
\hline
	fiducial model&	$\mathrm{color\,code}$\\
\hline
$r=0.00\,,\chi=0$	&$\mathrm{yellow}$\\
	$r=0.05\,,\chi=0\,;1$	&$\mathrm{black}$\\
	$r=0.10\,,\chi=0\,;1$	&$\mathrm{red}$\\
	$r=0.50\,,\chi=0\,;1$	&$\mathrm{blue}$\\
	$r=1.00\,,\chi=0\,;1$	&$\mathrm{green}$\\
\hline
\end{tabular}
\caption{Fiducial models used for forecasts. The remaining cosmological parameters are set to the Planck TT,TE,EE+lowTEB $\Lambda$CDM+r best-fit model, while $n_T=0$. For each value of the tensor-to-scalar ratio $r$, we consider the two cases of $\chi=0$ (no chirality) and $\chi=1$ (maximal chirality). Note that we just consider the case $\chi=0$ when $r=0$. The color code refers to the figures showing the posterior probability of $\chi$ in the next section.}\label{tab:fiducial}
\end{center}
\end{table}


\section{Results}
\label{sec:results}
In this section, we report our results in terms of both current limits from already existing data 
and forecasted sensitivity provided several fiducial models.
\subsection{Current limits}
\begin{figure}
\begin{center}
\includegraphics[width=0.9\textwidth]{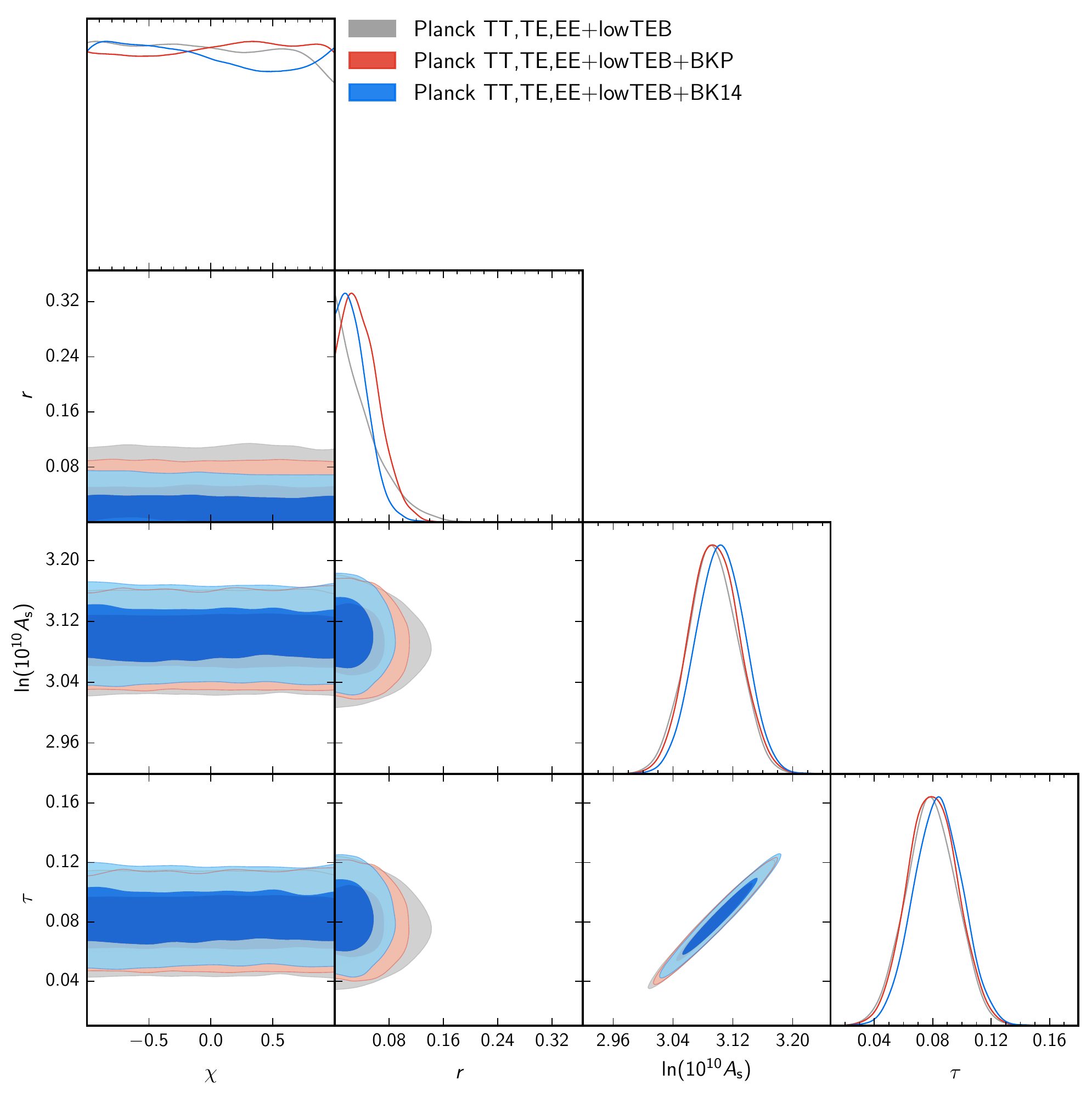}
\end{center}
\caption{Triangle plot showing the one-dimensional posterior distribution of the main parameter impacting the large scale CMB signal (top panel in each column) and their two-dimensional probability contours at 68\% and 95\% CL, for the indicated datasets.}\label{fig:tri}
\end{figure}

Figure \ref{fig:tri} and Table \ref{tab:current} show our results for the \lcdmrc  model for \textit{Planck} TT,TE,EE+lowTEB alone 
and in combination with data from Bicep/Keck, either BKP or the most recent BK14. 
As we can see, \textit{Planck} TT,TE,EE+lowTEB alone provides the weakest constraints on the model, 
given the low sensitivity of the large scale data alone to the tensor signal. The inclusion of BKP and 
BK14 data helps constrain better the tensor-to-scalar ratio, yielding $r<0.09$ and $r<0.07$ at 95\% CL 
respectively. 
In all cases, the chirality $\chi$ is left unconstrained. Given the definition of $\chi$ and the dependence of EB and TB spectra 
on the combination $r\cdot\chi$, one would have expected a degeneracy between the two parameters 
to arise. This is not the case here and we argue that it is partly due to the fact that the tensor signal is compatible with zero at high significance and that we do not have enough power 
to constrain $\chi$. 


When we include BK data, we are dramatically limiting the parameter space by better constraining $r$. As a result,
the inclusion of chirality only produces a slight broadening of the posterior of $r$ with respect to the bounds
reported in \cite{Ade:2015tva, Array:2015xqh}. The chirality parameter $\chi$ is still unconstrained, due to the fact that the tensor-to-scalar ratio is highly compatible with zero.

Finally, the inclusion of small scale data is crucial for constraining the amplitude of scalar perturbation, dramatically reducing the degeneracy between \loga and $r$.

\begin{table}
\begin{center}
\begin{tabular}{cccc}
\hline
\multirow{2}{*}{Parameter}	&Planck TT,TE,EE	&Planck TT,TE,EE	&Planck TT,TE,EE\\
&	+lowTEB&	+lowTEB+BKP&	+lowTEB+BK14\\
\hline\hline
\loga	&$3.093^{+0.063}_{-0.064}$	&$3.095^{+0.064}_{-0.063}$	&$3.103^{+0.062}_{-0.063}$\\
$\tau$	&$0.079^{+0.032}_{-0.033}$	&$0.080^{+0.033}_{-0.032}$	&$0.084^{+0.033}_{-0.032}$\\
$r$	&$<0.11$	&$<0.09$	&$<0.07$\\
$\chi$	&$unc.$	&$unc.$	&$unc.$\\
\hline
\end{tabular}
\caption{Constraints on the main parameters impacting the large scale CMB signal for the indicated datasets. The remaining cosmological parameters are fixed to the Planck TT,TE,EE+lowTEB bestfit for the \lcdm+r model. Limits are 95\% CL.}\label{tab:current}
\end{center}
\end{table}

\subsection{Forecasts for a COrE+ like mission}
We report here our forecasts for a future COrE+ like mission. 
We first consider a set of fiducial models with $\chi=0$ and then turn to analyze the same models 
with $\chi=1$. Apart from the tensor-to-scalar ratio $r$, which is set accordingly to Tab.\eqref{tab:fiducial}, 
the other cosmological parameters are always chosen to match the \textit{Planck} TT,TE,EE+lowTEB bestfit for 
the \lcdm+r model.

\subsubsection{Fiducial models with $\chi=0$}
\begin{figure}
\begin{center}
\includegraphics[width=0.5\textwidth]{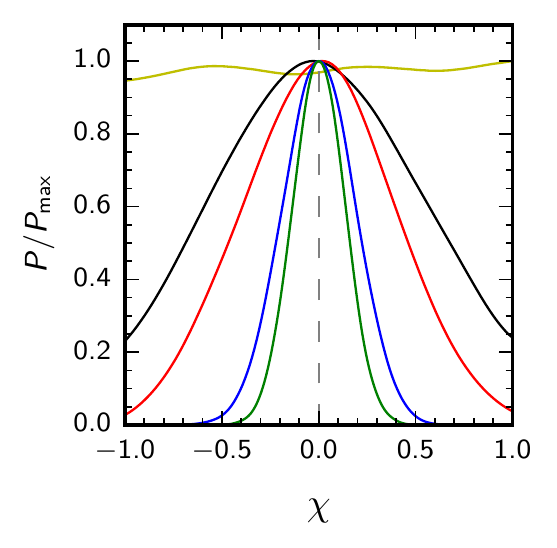}\includegraphics[width=0.5\textwidth]{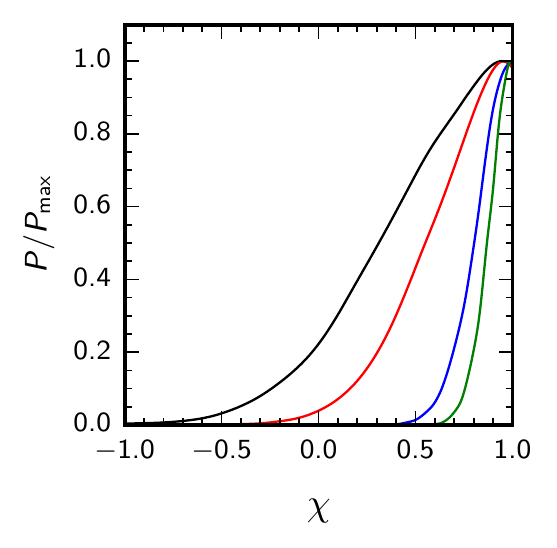}
\end{center}
\caption{One-dimensional posterior probability for the parameter $\chi$ for the forecasted sensitivity of a COrE+-like experiment for different fiducial models. The left panel refers to fiducial choice of $\chi=0$, while the right panel to $\chi=1$. The color code is the following: $r=0$ yellow (only in left panel), $r=0.05$ black, $r=0.1$ red, $r=0.5$ blue, $r=1.0$ green (see also Table \ref{tab:fiducial}). The vertical dashed line in the left panel corresponds to the fiducial value for $\chi=0$.}\label{fig:chi0}
\end{figure}

In Table \ref{tab:chi0}, we report limits on $r$ and $\chi$ for different fiducial models, 
while in the left panel of Fig.\ref{fig:chi0}, we show the one-dimensional posterior probability of the parameter $\chi$ 
for the same choice of models. As already mentioned, the sensitivity on $\chi$ depends strongly 
on the amplitude of primordial tensor perturbations. As we can see from the left panel of Fig.\ref{fig:chi0}, 
we can start constraining $\chi=0$ at 95\% CL for those models with $r>0.05$: assuming a fiducial 
$r=0.10$, which is the current 95\% upper limit on $r$, we recover $\chi=0.00^{+0.73}_{-0.72}$ at 95\% CL. 
For higher values of $r$, we find an increasing constraining power on $\chi$, with $\chi=0.00\pm0.26$ 
at 95\% CL for $r=1.0$. However, we stress that these cases with large $r$ are already excluded with 
high statistical significance by current CMB data. 

\begin{table}
\begin{center}
\begin{tabular}{cccccc}
\hline
Parameter	&$\chi=0,\,r=0$	&$\chi=0,\,r=0.05$	&$\chi=0,\,r=0.1$	&$\chi=0,\,r=0.5$	&$\chi=0,\,r=1.0$\\
\hline\hline
$r$	&$<2.4\cdot10^{-4}$	&$0.0500^{+0.0023}_{-0.0023}$	&$0.1001\pm0.0035$	&$0.500\pm0.011$	&$1.000\pm0.019$\\
$\chi$	&$unc.$	&$unc.$	&$0.00^{+0.73}_{-0.72}$		&$0.00^{+0.35}_{-0.35}$	&$0.00\pm0.26$\\
\hline
\end{tabular}
\caption{Constraints on the tensor-to-scalar ratio and the chirality parameter for the indicated fiducial models and a COrE+-like experiment. Limits are 95\% CL.}\label{tab:chi0}
\end{center}
\end{table}

\subsubsection{Fiducial models with $\chi=1$}
In Table \ref{tab:chi1}, we report limits on $r$ and $\chi$ for different fiducial models, 
while in the right panel of Fig.\ref{fig:chi0}, we show the one-dimensional posterior probability of the parameter $\chi$ 
for the same choice of models. Note that for this class of fiducial models, we do not consider the case 
$r=0$ for obvious reasons. As we can see from the right panel of Fig.\ref{fig:chi0}, we can start excluding $\chi=0$ 
at 95\% CL for those models with $r>0.05$: assuming a fiducial $r=0.10$, we get $\chi>0.235$ 
at 95\% CL. Even in this case, the bounds on $\chi$ becomes tighter for higher fiducial values of $r$. 
However, considering current limits on the tensor-to-scalar ratio and assuming a maximal parity-violating 
scenario (i.e., $\chi=1$), we expect that a future COrE+ like mission would be able to exclude $\chi=0$ 
at no better than $2.5\sigma$ roughly. For example, taking $r=0.05$ as our fiducial model, the lower bound 
at 95\% CL on $\chi$ is $\chi>-0.11$, with $\chi=0$ excluded at nearly $1.7\sigma$.


\begin{table}
\begin{center}
\begin{tabular}{ccccc}
\hline
Parameter	&$\chi=1,\,r=0.05$	&$\chi=1,\,r=0.1$	&$\chi=1,\,r=0.5$	&$\chi=1,\,r=1.0$\\
\hline\hline
$r$	&$0.0500\pm0.0023$	&$0.1000\pm0.0035$	&$0.500\pm0.011$	&$1.000\pm0.019$\\
$\chi$	&$> -0.11$	&$>0.24$	&$> 0.67$	&$> 0.78$\\
\hline
\end{tabular}
\caption{Constraints on the tensor-to-scalar ratio and the chirality parameter for the indicated fiducial models and a COrE+-like experiment. Limits are 95\% CL.}\label{tab:chi1}
\end{center}
\end{table}

We would like to discuss the possible implications that the recent results from \cite{Aghanim:2016yuo} in terms of a lower value of the optical depth $\tau$ could have on the analysis reported above. The dependence of non-vanishing TB and EB signal on the optical depth is mainly related to the reionization bump, \textit{i.e.} to the $\ell<20$ multipole region, where we expect the greatest contribution to chirality. A lower value of $\tau$ would reflect in less power at large scales, resulting in slightly broader constraints on $\chi$. Indeed, we have checked that this is the case by assuming a fiducial value of $\tau=0.06$ (instead of the fiducial value $\tau=0.079$ adopted previously) and performing again forecasts. We find that limits on $\chi$ broadens roughly by a factor of $\sim 0.1\sigma$. As an example, for the fiducial model with $r=0.1$ and $\chi=0$ ($\chi=1$), we get a 95\% CL on the chirality parameter of $\chi=0.00\pm0.78$ ($\chi>0.14$), to be compared with the equivalent bounds in Tab.\ref{tab:chi0} (Tab.\ref{tab:chi1}).

Before concluding this section, we remind that cosmological birefringence models predict a non-vanishing TB and EB signal as well, thus introducing some level of degeneracy with chiral gravitational waves. However, as thoroughly discussed in \cite{Gluscevic:2010vv}, the two effects are almost orthogonal, with chirality being a pure tensor contribution, thus mostly affecting large scales and being dumped at smaller scales.

\section{Conclusions}
\label{sec:conclusions}


We have discussed the sensitivity of current (Planck and Bicep/Keck) and future (COrE+) CMB experiments to the chirality of primordial gravitational waves $\chi$, employing the full set of temperature and polarization power spectra. Our main conclusion is that unfortunately the two-point correlation function currently used to build CMB likelihood, can only weakly constrain chirality. This is due to fact that $\chi$ can be constrained only if the amplitude of the primordial tensor signal $r$ is detected to be different from zero and high enough to induce a detectable signature in the parity violating spectra $TB$ and $EB$, which is not the case. Current power spectrum datasets are totally insensitive to chirality models, as shown in Fig.\eqref{fig:tri}. The performed forecast for $\chi$ with a COrE+ like experiment are given in Fig.\eqref{fig:chi0}, assuming fiducial models with no parity violation ($\chi=0$, left panel) and maximal parity violation ($\chi=1$, right panel), respectively. The most stringent constraint on $\chi$ assuming a fiducial value $r=0.1$ (still marginally in agreement with current bounds on the amplitude of tensor modes) is $\chi=0.00^{+0.73}_{-0.72}$ and $\chi>0.24$ at 95\% CL respectively. In other words, in the best case scenario of perfect foreground removal and high tensor-to-scalar ratio roughly compatible with current limits, we could be able to constrain chirality models at $\sim2.5\sigma$ at best.
We stress that when we resctrict to the low-$\ell$ (where chiral gravity models predict most of the signal) \textit{Planck} measurements, we employ only LFI data to derive the constraints on $\chi$\footnote{We briefly comment about the future possibility to consider HFI data at low $\ell$: even if it would be interesting to investigate it, we do not expect great improvements on the detectability of $\chi$. The reason can be easily drawn from the analysis for the COrE+ like experiment and relies on the fact that the current constraints on $r$ are already limiting the room for $r$ different from zero.}.
In the ideal cosmic variance limited scenario, with full sky coverage, a Fisher matrix approach predicts roughly a $5\sigma$ detection of $\chi$ for $r=0.1$, which, according to the Cram\'er-Rao inequality, can be considered as a lower limit on achievable sensitivity. Such a neat detection (i.e. $5\sigma$) would be possible only in a very optimistic scenario (perfect foreground removal and $f_\mathrm{sky}=1$). On the other hand, from the same analysis, we find that for values of $r\leq0.008$ chirality turns out be undetectable even in the best-case cosmic-variance-limited scenario. 

In conclusion, our results suggest that the two-point correlation function is not the right tool to constrain a nearly scale-invariant $\chi$ even for high-precision future CMB experiments. However, the results should be sensitive to the scale dependence of $\chi$, and TB and EB will be informative if the GW power spectrum has a nontrivial peak on large scales \cite{Namba:2015gja}.
Moreover, tests of higher order statistics \footnote{See \cite{Maldacena:2011nz, Soda:2011am, Shiraishi:2011st, Zhu:2013fja, Cook:2013xea} for studies on parity-violating GW NG.} including clean information on parity violation, such as odd (even) $\ell_1 + \ell_2 + \ell_3$ of TTT, TTE, TEE, TBB, EEE and EBB (TTB, TEB, EEB and BBB) \cite{Kamionkowski:2010rb,Shiraishi:2011st} in principle, enhance the detectability of the chirality of GWs \cite{Shiraishi:2012sn,Shiraishi:2013kxa,Shiraishi:2014roa,Shiraishi:2014ila,Ade:2015ava,Namba:2015gja}. 
We leave such interesting topics for future investigation.

\acknowledgments
We are grateful to Luca Pagano for precious help with Boltzmann code and forecasts and to Giovanni Cabass, Massimiliano Lattanzi and Sabino Matarrese for useful discussions. 
This paper is based on observations obtained with the satellite {\sc Planck} 
(http://www.esa.int/Planck), an ESA science mission with instruments and contributions directly funded by ESA Member States, NASA, and Canada. We acknowledge the support by ASI/INAF Agreement 2014-024-R.1 for the Planck LFI Activity of Phase E2.
We acknowledge the use of computing facilities at NERSC (USA), of the HEALPix package [23], and of the Planck Legacy Archive (PLA). 
MG was partly supported by the grant ``Avvio alla ricerca''  for young researchers by ``Sapienza'' university and is supported by the Vetenskapsr\aa det (Swedish Research Council). 
MS was supported in part by a Grant-in-Aid for JSPS Research under Grants No.~27-10917, and in part by the World Premier International Research Center Initiative (WPI Initiative), MEXT, Japan.


\appendix
\section{Fisher matrix computations} \label{appen:fish}

\begin{figure}
\begin{center}
\includegraphics[width=0.9\textwidth]{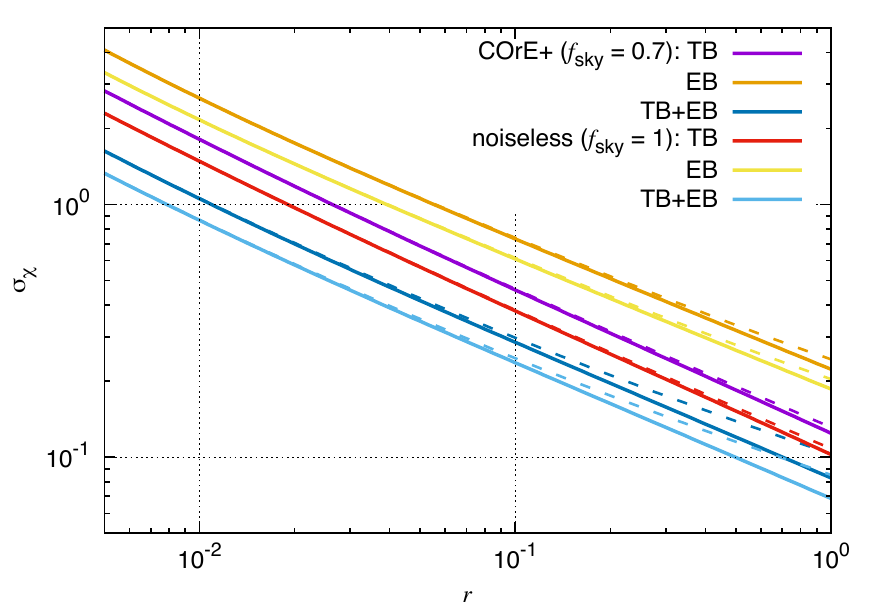}
\end{center}
\caption{Expected $1\sigma$ errors on $\chi$ ($\sigma_\chi$) as a function of $r$, computed from the Fisher matrix \eqref{eq:Fisher_TB+EB}. We here analyze the sensitivities in a COrE+-like experiment and an ideal full-sky noiseless experiment. Solid (dashed) lines correspond to the results for $\chi = 0$ ($1$).}\label{fig:error_chi_COrEplus}
\end{figure} 

In this appendix, we estimate the sensitivities to the chirality parameter $\chi$ in a simple way and 
compare them with the results from a Monte Carlo analysis discussed in Sec.~\ref{sec:results}.

We here consider the measurements of $\chi$ using only TB and EB correlations. 
Assuming that off-diagonal components of the covariace matrix are negligibly small compared with 
diagonal ones, the Fisher matrix for $\chi$ is expressed as \cite{Gluscevic:2010vv}
\begin{eqnarray}
  F = f_{\rm sky} \sum_{\ell = 2}^{\ell_{\rm max}}
  \frac{\partial {\bf C}_\ell}{\partial \chi} {\bf Cov}_\ell^{-1} \frac{\partial {\bf C}_\ell^\top}{\partial \chi} ~, \label{eq:Fisher_TB+EB}
  \end{eqnarray}
where $ \partial {\bf C}_\ell / \partial \chi \equiv (C_\ell^{TB}|_{\chi = 1}, C_\ell^{EB}|_{\chi = 1})$ and 
\begin{eqnarray}
  {\bf Cov}_\ell = \frac{1}{2\ell + 1}
  \left(
  \begin{array}{cc}
    \tilde{C}_\ell^{TT} \tilde{C}_\ell^{BB} + (\tilde{C}_\ell^{TB})^2 &
    \tilde{C}_\ell^{TE} \tilde{C}_\ell^{BB} + \tilde{C}_\ell^{TB} \tilde{C}_\ell^{EB}  \\
    \tilde{C}_\ell^{TE} \tilde{C}_\ell^{BB} + \tilde{C}_\ell^{TB} \tilde{C}_\ell^{EB} &
    \tilde{C}_\ell^{EE} \tilde{C}_\ell^{BB} + (\tilde{C}_\ell^{EB})^2 
  \end{array}
 \right) ~. \label{eq:cov_TB+EB}
  \end{eqnarray}
with $\tilde{C}_\ell$ denoting the sum of the primordial signal $C_\ell$, additional signal produced via 
gravitational lensing and instrumental noise spectrum. We assume that the noise spectra of 
TE, TB and EB are zero, and drop negligibly small contributions of lensed TB and EB modes \cite{Ferte:2014gja}. Expected $1\sigma$ error on $\chi$ is given by $\sigma_\chi = 1 / \sqrt{F}$. 

Figure~\ref{fig:error_chi_COrEplus} describes our numerical results of $\sigma_\chi$ in a 
COrE+ like measurement and an ideal noiseless full-sky measurement, showing that the sensitivity 
to $\chi$ gets worse as $r$ becomes small due to the decrease of ${\bf C}_\ell (\propto r)$. 
Because of this feature, for $r \lesssim 0.01$, $\chi$ is undetectable, even in an ideal noiseless full-sky 
measurement. As seen in this figure, $\sigma_\chi$ depends very weakly on $\chi$ for small $r$, 
since the contributions of $\tilde{C}_\ell^{TB}$ and $\tilde{C}_\ell^{EB}$ to the covariance 
\eqref{eq:cov_TB+EB} then become subdominant. The results for a COrE+-like survey in 
Fig.~\ref{fig:error_chi_COrEplus} are almost consistent with the results obtained via a full Monte Carlo 
analysis in Sec.~\ref{sec:results}, although we here compute the Fisher matrix \eqref{eq:Fisher_TB+EB} 
by fixing the cosmological parameters other than $\chi$ and $r$ and this leads to a bit better sensitivity. Also, 
we remind that Fisher matrix results can be considered as a lower bound on the variance, 
according to the Cram\'er-Rao inequality.

\end{document}